\documentclass[twocolumn]{aastex631}
\usepackage{amsmath}
\usepackage{graphicx}
\usepackage{amssymb}
\usepackage{array}
\usepackage[varg]{txfonts}
\usepackage{color}
\usepackage[utf8]{inputenc}
\usepackage{comment}
\usepackage{ulem}

\makeatletter
\providecommand*{\diff}%
	{\@ifnextchar^{\DIfF}{\DIfF^{}}}
\def\DIfF^#1{%
	\mathop{\mathrm{\mathstrut d}}%
		\nolimits^{#1}\gobblespace}
\def\gobblespace{%
	\futurelet\diffarg\opspace}
\def\opspace{%
	\let\DiffSpace\!%
	\ifx\diffarg(%
		\let\DiffSpace\relax
	\else
		\ifx\diffarg[%
			\let\DiffSpace\relax
		\else
			\ifx\diffarg\{%
				\let\DiffSpace\relax
			\fi\fi\fi\DiffSpace}
\makeatother

\newcommand{\beqa}{\begin{eqnarray}}
\newcommand{\eeqa}{\end{eqnarray}}
\newcommand{\ba}{\begin{array}}
\newcommand{\ea}{\end{array}}
\def\d{\text{d}}

\def\({\left(}
\def\){\right)}

\begin{document}
\title{Tidally-induced Magnetar Super Flare at the Eve of Coalescence with Its Compact Companion}

\author[0000-0003-4673-773X]{Zhen Zhang}
\affiliation{Key Laboratory of Particle Astrophysics, Institute of High Energy Physics, Chinese Academy of Sciences, \\
19B Yuquan Road, Beijing 100049, People’s Republic of China}

\author[0000-0001-7599-0174]{Shu-Xu Yi}
\affiliation{Key Laboratory of Particle Astrophysics, Institute of High Energy Physics, Chinese Academy of Sciences, \\
19B Yuquan Road, Beijing 100049, People’s Republic of China}

\author[0000-0001-5586-1017]{Shuang-Nan Zhang}
\affiliation{Key Laboratory of Particle Astrophysics, Institute of High Energy Physics, Chinese Academy of Sciences, \\
19B Yuquan Road, Beijing 100049, People’s Republic of China}
\affiliation{University of Chinese Academy of Sciences, Chinese Academy of Sciences, Beijing 100049, China}

\author[0000-0002-4771-7653]{Shao-Lin Xiong}
\affiliation{Key Laboratory of Particle Astrophysics, Institute of High Energy Physics, Chinese Academy of Sciences, \\
19B Yuquan Road, Beijing 100049, People’s Republic of China}

\author[0000-0003-2957-2806]{Shuo Xiao}
\affil{Guizhou Provincial Key Laboratory of Radio Astronomy and Data Processing, Guizhou Normal University, Guiyang 550001, People’s Republic of China}
\affil{School of Physics and Electronic Science, Guizhou Normal University, Guiyang 550001, People’s Republic of China}

\correspondingauthor{Zhen Zhang, Shu-Xu Yi}
\email{zhangzhen@ihep.ac.cn, sxyi@ihep.ac.cn}

\begin{abstract}
In the late inspiral phase of a double neutron star (NS) or NS-black hole system in which one NS is a magnetar, the tidal force on the magnetar arisen from its companion will increase dramatically as the binary approaches. The tidal-induced deformation may surpass the maximum that the magnetar's crust can sustain just seconds or subseconds before the coalescence. A catastrophic global crust destruction may thus occur, and the magnetic energy stored in the magnetar's interior will have the opportunity to be released, which would be observed as a {\it superflare} with energy 100s of times larger than giant flares of magnetars.  Such a mechanism can naturally explain the recently observed precursor of GRB 211211A, including its quasiperiodic oscillation. We predict that in the coming gravitational wave O4/O5 period, there could be a fraction of detected double NS mergers associated with such super flares. If observed, copious information on the structure and magnetic field in an NS interior can be obtained, which is hard to study elsewhere.  
\end{abstract}


\section{Introduction}

The discovery of GW170817 and the follow-up multi-messenger observations in $\gamma$-ray, X-ray, optical, and radio link $\gamma$-ray bursts (GRBs), kilonovae (KNe) and gravitational waves (GWs), and pinpoint its origin to a double neutron star (DNS) merger \citep{NAKAR20201,doi:10.1146/annurev-astro-112420-030742}. The prompt $\gamma$-ray emission (GRB 170817A) was detected $\sim$1.7\,s after the merger, suggesting that it spent $\sim1$ s to break out from the circumburst medium \citep{Gottlieb:2022sis}. Although GW170817, GRB 170817A, and AT 2017gfo remain the unique event which was observed as GWs, GRBs and KNe, there have been several GRBs that are found to be associated with KNe, and are thought to have originated from DNS mergers, \textit{e.g.,} GRB 130603B \citep{2013Natur.500..547T}.

GRB 211211A is a recently detected GRB, which was associated with an AT 2017gfo-like KN \citep{Rastinejad:2022zbg, Xiao:2022quv}. No associated supernova was observed. Its main burst (MB) behaved like a long GRB, lasting for $T_{90}\sim34.4$\,s. The isotropic equivalent energy is $E_{{\rm iso}}\sim5.3\times10^{51}$ erg in its MB, with a beam-corrected total energy of $\sim6.6\times10^{48}$ erg. Its optical counterpart, together with the constraint on the emitting region size and its specific positions in empirical correlation relationships suggest that the GRB should be classified into the merger-originated category \citep{Xiao:2022quv}, which commonly have $T_{90}\lesssim2\,$s.

Interestingly, GRB 211211A was found with a precursor: a $\sim0.2$\,s duration flare $\sim1$\,s prior to its MB \citep{Xiao:2022quv}, 
where the separation of $\sim1$ s is roughly consistent with the time lag of GRB170817A to the merger. 
The isotropic energy of the precursor is estimated at $\sim7.7\times10^{48}$\,erg, comparable to the beam-corrected energy of the MB. 
Here the beaming of the MB is due to the relativistic jet launched after coalescence \citep{Gottlieb:2022sis}. Since the precursor is produced before coalescence through an independent mechanism, it may not be necessarily beamed \citep{Tsang2013,Metzger2016}.
More intriguingly, the precursor showed a quasiperiodic oscillation (QPO) at $\sim22.5$ Hz \citep{Xiao:2022quv}.

Various studies have been carried out for the precursors in short GRBs (sGRBs), e.g., \cite{2017PhRvD..95f3016C}, \cite{2013MNRAS.428..518D}, and \cite{2010ApJ...723.1711T}. 
There are two main branches of theoretical models interpreting $\gamma$-ray precursors in the merger-induced GRBs: One attributes the precursor to the magnetosphere interaction between two neutron stars (NSs) \citep{Vietri1996,Hansen2001Lyutikov,wang2018pre} or the resonant shattering of NS crusts \citep{palenzuela2013electromagnetic} before coalescence, and the other seeks the origin of the precursor in the newly born hypermassive magnetar after coalescence. The former has difficulty in explaining the huge energy release during the precursor and its $0.2~{\rm s}$ duration \citep{Tsang2013,Metzger2016}, while the latter needs to solve the problem of the large optical depth of the collision ejecta, which is thought to block photons at the timescale of seconds prior to the MB \citep{Gottlieb:2022sis}.

Here we suggest that the GRB 211211A precursor is related to a magnetar in a coalescing DNS system. Three giant flares (GFs) have been observed from magnetars \citep{Aptekar_2001,strohmayer2005discovery,watts2006detection}, which are believed to be powered by the global reconfiguration of the magnetar's magnetic field, as a result of the crust fracturing event \citep{Thompson95,Thompson2001,Watts:2006mr,doi:10.1146/annurev-astro-081915-023329}. Usually, flares or bursts from magnetars are believed to be roughly isotropic, or at least not obviously beamed \citep{Thompson95,Metzger2016,Xiao21}. The energy released in the GRB 211211A precursor is $\sim2$ orders of magnitude larger than that of GFs, hinting at the occurrence of a much more catastrophic event. Hereafter, we call such an event a {\it superflare} (SF), which may involve the release of magnetic energy interior to the magnetar.

Generally, the NS (inner) crust cannot be broken directly by the tidal force of the companion until the DNS merger. However, we find that if the crust is highly magnetized, it can be disrupted before the coalescence. In this work, we present a mechanism that attributes the precursor to the catastrophic magnetar SF, which is the primary mechanism for excavating the interior magnetic energy through the tidal-induced breaking of the magnetized crust prior to coalescence. First, we examine this scenario by comparing the tidal force on the magnetar raised by the companion to the maximum that the magnetized crust can sustain. Then we estimate the energy release of the SF, its duration and the possible QPO origin.

\section{Tidal-induced Crust Breaking}
\label{sec:1} 

An NS crust is usually modeled as body-centered cubic (bcc) Coulomb crystals composed of identical ions \citep{HANSEN20041,Baiko:2017hnf}. The stress tensor in the crust can be expressed as \citep{Baiko:2018jax}:
\begin{equation}
    \tilde{\sigma}_{ij}=\frac{n\,Z^2\,e^2}{a}\sigma_{ij},
\end{equation}
where $n$ is the ion number density, $Z$ is the charge number, $a=(4\pi\,n/3)^{-1/3}$ ion-sphere radius, and $\sigma_{ij}$ is the dimensionless stress tensor. The lattice is assumed to be in the ground-state under its own gravity of the NS, i.e., a nondeformed bcc lattice with $\sigma_{ij}=-\zeta\delta_{ij}$, where $\zeta=0.8959293$ is the Madelung constant. The crust is crystallized in hydrostatic state, where the pressure it feels is isotropic.

If an external anisotropic stress exerts on the crust, the lattice will be deformed, and the stress tensor $\sigma_{ij}$ will no longer be isotropic (i.e., $\sigma_{ij}\ne\sigma\,\delta_{ij}$) \citep{Baiko:2017hnf,Baiko:2018jax}. The excess energy density due to the deformation is:
\begin{equation}
    \Delta\tilde{P}=\frac{n\,Z^2\,e^2}{a}\Delta P,
\end{equation}
where $\Delta P$ is the pressure anisotropy in units of $Z^2\,e^2/a$. The excess pressure anisotropy corresponds to the fractional length change $\epsilon\equiv\delta L/L$. As found in \cite{Baiko:2017hnf}, the critical (dimensionless) excess energy density, beyond which the crust will break, is $\Delta P_{\rm{cri}}\sim0.01-0.04$, and the corresponding $\epsilon_{\rm{cri}}\sim0.03$. 

Now we consider a plate of crust with area $S$ and original thickness $l$. The plate is elongated along the thickness direction (vertical to the plate) to the critical extent $\epsilon$. The excess energy in this plate is
\begin{equation}\label{tildeU}
    \Delta\tilde{U}=S\,l\,\Delta\tilde{P}.
\end{equation}
Therefore, the elastic restoring force is
\begin{equation}\label{eq:Fela}
    F_{\rm{ela}}=\frac{\Delta\tilde{U}}{\epsilon\, l}.
\end{equation}
The maximum elastic restoring force corresponds to $\Delta P_{\rm{cri}}$ and $\epsilon_{\rm{cri}}$. Substituting them into Eq.~\eqref{eq:Fela}, one gets the maximum deforming external force $F_{\rm{ext,max}}$ that the crust can sustain, namely
\begin{equation}
    F_{\rm{ext,max}}=S\,\frac{\,n\,Z^2\,e^2\,}{a}\,\frac{\Delta P_{\rm{cri}}}{\epsilon_{\rm{cri}}}.
\label{eqn: Fext}
\end{equation}
Generally, the ionic Coulomb crystals comprising a crust can be deformed if there is a nonuniform magnetic field \citep{Baiko:2017hnf,Baiko:2018jax} in the crust. Therefore, in Equation (\ref{eqn: Fext}), we should also include the contribution of the nonuniform magnetic field by replacing $n$ (and the resulting $a$) with $n_{\rm{eff}}=(1-b)\,n$ (see the Appendix for physical details).
We require $0<b<1$, otherwise the crust could not remain stable. The parameter $b$ can be related with the crustal magnetic field as
\begin{equation}
 B\,\delta\,\!B\sim4\pi\,\Delta\tilde{P}\,\left[1-\(1-b\)^{4/3}\right],
 \label{BdB}
\end{equation}
where $\delta B$ is the nonuniform part of the crustal magnetic field strength. 
Here, $\delta\,\!B$ can have two likely contributions: One comes from the evolution of the crustal seed magnetic field, and the other originates from the tidal-torque-induced magnetic field amplification during the inspiral phase \citep{2013MNRAS.428..518D}. The latter may lead to crust breaking, referred to as the Dall' Osso-Rossi (DR) mechanism, due to the excess of amplified magnetic stress. In reality, the magnetic stress will always be assisted by the tidal stress, making the crust disruption earlier and easier than expected with a pure DR mechanism.
Equation (\ref{BdB}) leads to:
\begin{equation}\label{BBlimit}
 B\gtrsim\sqrt{4\pi\,\Delta\tilde{P}\,\left[1-\(\frac{\rho_{{\rm eff}}}{\rho}\)^{4/3}\right]},
\end{equation}
where $\rho_{{\rm eff}}=(1-b)\,\rho$, and $\rho$ is the mass density of the crust. In the above inequality, $B\gtrsim\sqrt{B\,\delta B}$ is applied, because $B\gtrsim\delta B$ is generally expected. We will employ the above equation to give a rough estimation of the crustal magnetic field strength later.

At the vicinity of coalescence of the binary, the crust of the magnetar that faces and is opposite the companion feels an elongating tidal force:
\begin{equation}
    F_t\sim\frac{~GM_c~}{R^2}\frac{~l~}{R},
\end{equation}
where $M_c$ is the mass of the companion star, and $R$ is the binary orbital separation.

We define the ratio between $F_t$ and $F_{\rm{ela,max}}$ as $\eta$, which is a function of the binary separation $R$:
\begin{equation}
    \eta=\frac{G\,M_c\,l\,a_{{\rm eff}}\,\epsilon_{\rm{cri}}}{R^3\,\Delta P_{\rm{cri}}\,n_{{\rm eff}}\,Z^2\,e^2\,S}.
    \label{eqn:eta}
\end{equation}
In the inspiral phase, $R$ shrinks due to the GW radiation:
\begin{equation}
    R=R_0\,\(\frac{t_c-t}{t_0}\)^{1/4},
    \label{eqn:R}
\end{equation}
where $t_c$ is the instance of coalescence, and $R_0$ is the binary separation at the reference time $t_0$. 
A circular orbit is assumed in the above, as the orbit should be long circularized due to GW emission. 
This equation describes the point-mass approximation properly. It assumes that coalescence occurs when $t=t_{c}$, i.e., at $R=0$. However, two NSs touch each other when $R$ equals the sum of their tidally elongated radii, i.e., at $R\sim2.4~R_{\rm NS}$ if the mass ratio is unity \citep{2013MNRAS.428..518D}, where $R_{\rm NS}$ denotes the original NS radius. Take a typical value, $R_{\rm NS}=10 ~{\rm km}$; thus, the merger occurs at $R\sim24~{\rm km}$, which introduces a correction of $\sim0.001~{\rm s}$ to the merger time. In a more realistic treatment, tidal deformation of finite-sized NSs can cause a deviation from the point-mass potential, whereas this deviation is on the order of $\sim0.1\%-1\%$ or less before the merger \citep{PhysRevD.85.123007}. So the correction to the orbital evolution should be $\lesssim1\%$. Therefore, such effects are negligible for our purpose.
Accordingly, we rewrite the tidal force as function of time:
\begin{equation}
\label{eq:tidalFt}
F_{t}\sim\frac{G\,M_{{\rm c}}}{R_{{\rm 0}}^2}\frac{l}{R_{{\rm 0}}}\(\frac{t_{c}-t}{t_{{\rm 0}}}\)^{-3/4}. 
\end{equation}

For an equal-mass DNS system with mass $M_{1,2}=1.4\,M_\odot$ \citep{Rastinejad:2022zbg}, $R_0=229\,$km at $t_0=10$\,s, where $M_\odot$ is the solar-mass. 
 Therefore, $\eta$ as function of time is:
\begin{eqnarray}
\label{eta}
\eta=&0.8\,l_{1}\,S^{-1}_{\!100}\,\epsilon_{{\rm cri},0.03}\,\Delta P^{-1}_{{\rm cri},0.014}\,Z^{-2}_{38}\,A^{4/3}_{300} \nonumber\\
&\times\,\rho^{-4/3}_{{\rm eff},11}\,M_{c,1.4}\,\(\frac{t_c-t}{1\,{\rm s}}\)^{-3/4},
\hspace*{10mm}
\end{eqnarray}
where $\rho_{\rm{eff},11}=\(1-b\)\rho_{11}$ is the effective density in units of $10^{11}$\,g/cm$^3$, $l_1$ is the height of the destructed crust in units of 1 km, $M_{c,1.4},~\epsilon_{\rm{cri},0.03},~\Delta P_{\rm{cri},0.014},~Z_{38},~A_{300}$ (ion mass number), and $S_{100}$ are the corresponding quantities scaled with 1.4 $M_{\odot}$, 0.03, 0.014, 38, 300, and 100 km$^2$, respectively.

\begin{figure}
\centerline{
\includegraphics[width=\columnwidth]{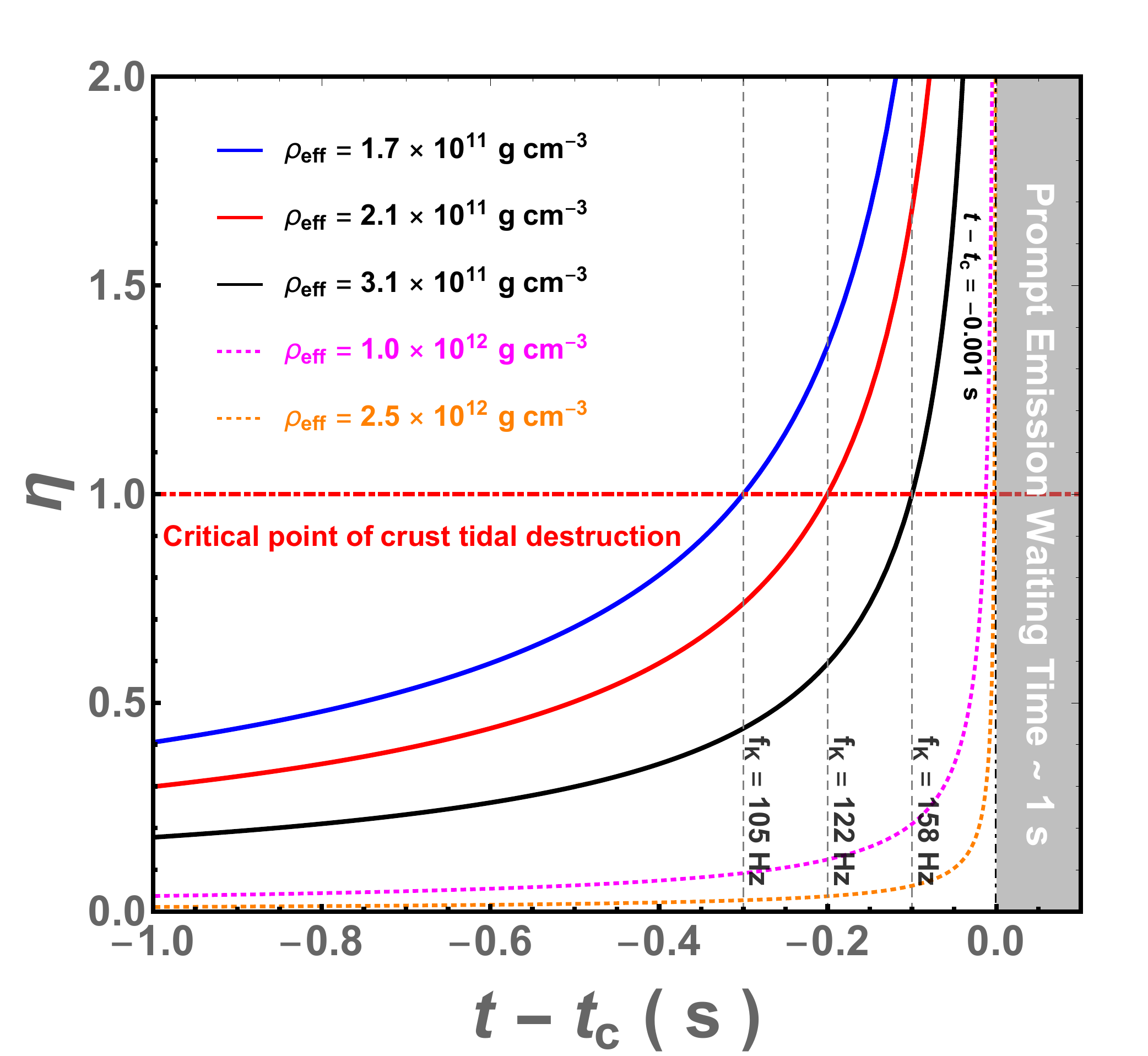}
}
\caption{The ratio $\eta$ between the tidal force arisen from the companion and the critical distorting force that the crust can sustain, as a function of time; $t_c$ is the coalescence instance. When $\eta$ surpasses unity, the crust is thought to be destructed. Different colors correspond to different $\rho_{\rm{eff}}$, and the orbital frequencies at the critical points are labeled.}
\label{fig:Sfactor}
\end{figure}

In Fig.~\ref{fig:Sfactor}, we plot $\eta$ as function of $t$; $\eta\(t\)$ surpasses unity when the tidal force from the companion becomes strong enough to crack the crust. 
In Fig.~\ref{fig:Sfactor}, $l_1=1$ when the creaking extends to the inner crust, $\epsilon_{\rm{cri},0.03}=1$, $\Delta P_{\rm{cri},0.014}=1$, $A_{300}=1$, and $Z_{38}=1$ \citep{Piro_2005,Baiko:2018jax}, respectively. The destructive area $S$ is estimated for GRB 211211A as follows: the precursor's rising time is about $t_{{\rm r}}\sim6-10$ ms, which indicates that the corresponding size of the cracked crust is $v_{A}\,t_{{\rm r}}$, where  $v_A$ is the propagation speed of Alfv$\acute{\rm{e}}$n waves. For a typical magnetar with a magnetic field of $\sim10^{14}$ G, the destruction area is $S\sim\(v_{A}\,t_{{\rm r}}\)^2\gtrsim100 ~{\rm km}^{2}$. The estimated $S$ is in the same order of magnitude as the surface area of the magnetar, which indicates that the destruction is a global event. Thus, $S_{100}=1$. As demonstrated in Fig.~\ref{fig:Sfactor}, as $R$ shrinks, the tidal force increases sharply with time. For $\rho_{{\rm eff},11}=1.7,~2.1,~3.1$, $\eta\(t\)$ exceeds unity at time $t\sim-0.3,~-0.2,~-0.1$ s.
Thus a burst is produced $\sim0.2$ s before the merger and continues until the merger, which is consistent with the SF's duration. Then the two NSs touch each other, followed by the prompt emission waiting time of $\sim~1~{\rm s}$ \citep{Gottlieb:2022sis}, as shown in Fig.~\ref{fig:Sfactor} by the gray region. This indicates that $\rho_{{\rm eff}}\sim2.1\times10^{11}~{\rm g~cm^{-3}}$. If the intrinsic mass density of the inner crust is $\rho\gtrsim10^{13}~{\rm g~cm^{-3}}$, one has $B\gtrsim1.0\times10^{15}$ G inside the crust from Equation.~\eqref{BBlimit}.

\section{The Energy Resource of the Precursor and the Origin of Its QPO}
\label{sec:2}

The energy released from the crust breaking can be estimated with \citep{Baiko:2018jax}:
\begin{equation}
    Q=3.3\times10^{41}\,\rho^{4/3}_{11}\,Z^2_{38}\,A^{-4/3}_{300}\,l_1\,S_{\!100}\,~\rm{erg},
\end{equation}
where the mass density is about $\rho_{11}=10^{3}$ for the inner crust.
The maximum energy available through crust creaking is thus $\sim10^{45}$\,erg, when the entire outer and inner crust on the surface of the NS is destructed. However, the energy is still $\sim$4 orders of magnitude less than the observed energy release of $\sim7.7\times10^{48}$\,erg in the precursor.

Another energy reservoir is its strong magnetic field. The magnetic energy stored in the magnetosphere, $E_{\rm{sph}}$ can be evaluated by volumetrically integrating the dipole magnetic field energy density outside the NS.
With the surface magnetic field $B_{0}\sim2\times10^{15}$\, G, same as the strongest magnetic field of known magnetar SGR 1806-20 \citep{doi:10.1146/annurev-astro-081915-023329}, $E_{\rm{sph}}$ is calculated to be 
\begin{equation}
    E_{\rm{sph}}\sim\frac{1}{6} \,B_{0}^2\,R^3_{NS}\sim6.7\times10^{47}\,\rm{erg}.
\end{equation}
It is still about 1 order of magnitude lower than the observed precursor. Therefore, the energy in the magnetosphere is not enough even if depleted completely. 

The stored magnetic energy $E_{{\rm core}}$ in the interior magnetic field of a magnetar can be $1-2$ orders of magnitude higher than $E_{\rm{sph}}$ \citep{Link_2016}, up to $E_{{\rm core}}\sim10^{50-51}$\,erg \citep{Reisenegger2009,Reisenegger:2008yk}. We therefore argue that, the observed precursor involves the releasing of interior magnetic energy, during a global destruction of its crust. We also provide an effective way of tapping and releasing the interior magnetic field energy in the Appendix.

Next, we turn to the possible origin of the observed QPO in the precursor. As shown in Fig. \ref{fig:Sfactor}, the orbital frequency of the binary is $f_{\rm K}\sim100$ Hz, well above the observed $\sim22.5$ Hz. It means that the QPO is not associated with the orbital periodicity.   
A natural candidate is the spin periodicity of the magnetar. The spin period is independent of the orbital period since the NS is not tidally locked during the inspiral \citep{Cutler1992}. The magnetar spins down under magnetic dipole radiation at the rate: 
\begin{equation}
    \dot{P}\sim\frac{10^{-9}B_{0,15}^2}{P_1} ~{\rm s/s},
\end{equation}
where $B_{0,15}$ is the surface magnetic field strength in units of $10^{15}$\,G, and $P_1$ is the NS spin period scaled with 1\,s. The characteristic age of the magnetar is
\begin{equation}
    \tau=\frac{1}{2}\,\frac{P}{\dot{P}}\sim15\,\frac{P_1^2}{B^2_{0,15}}\,{\rm years}.
\end{equation}
If the observed QPO period $1/(22.5\,\rm{Hz})\approx0.04$\,s is the spin period, the characteristic age of the magnetar is less than half a year before the merger, which is very unlikely. In other words, it is impossible for the magnetar to remain at such a high spin frequency under its strong magnetic dipole radiation.

Magnetar QPOs were detected after the GF from SGR 1806-20 for the first time \citep{IsraelEt.al.2005}, with frequencies comparable to $\sim22.5$ Hz found in GRB 211211A, which were interpreted as the magnetoelastic/crustal oscillations in magnetar crusts or interiors \citep{IsraelEt.al.2005,Levin:2006ck,samuelsson2007neutron,gabler2011magneto,Link_2016,2022MNRAS.tmp.1632N,2022arXiv220511112S}.
Taking the crustal oscillations for example,
the magnetosphere attached to the quasiperiodically deformed crust could lead to periodic changes in the luminosity \citep{Gabler:2014bza}.
If so, the QPO frequency is mainly determined by the tension of the magnetic field and the bulk stellar properties.

Both the elastic crust and the liquid core can be treated as perfect conductors.  
These two parts are connected by magnetic fields.
The effective tension inside each part can be estimated as
\beqa
\label{eq:tension}
T_{{\rm eff}}\sim\frac{B^{2}_{{\rm eff}}}{4\pi},
\hspace*{10mm}
\eeqa 
where $B_{\rm eff}$ represents the effective magnetic field strength. Here, $B_{{\rm eff}}=B$ inside the crust, and $B_{{\rm eff}}=\sqrt{B\,B_{{\rm cri}}}$ inside the core, where $B$ is the strength of the local magnetic field, and $B_{{\rm cri}}$ is the strength of the magnetic field inside the superconducting flux tubes \citep{Levin:2006ck}.

The frequency of the crustal oscillation or the magnetoelastic oscillation of the $k$th mode $\nu_{k}$ can be roughly estimated using the simplified model proposed in \cite{Levin:2006ck} as 
\beqa
\label{eq:qpofre}
\nu_{k}\sim\frac{1}{2\pi}\sqrt{\(\frac{k\,\pi}{l_k}\sqrt{\frac{T_{\rm eff}}{\rho_{k}}}\)^2-\(\frac{1}{\tau_{k}}\)^2}, 
\hspace*{10mm}
\eeqa 
where $\rho_{k}$ denotes the local mass density involved in the $k$th mode oscillation. Here, $k$ is an integer labeling the oscillatory mode, $l_{k}$ denotes the characteristic scale on which the oscillations occur in a certain dimension, and $\tau_{k}$ is the corresponding oscillation dissipating timescale, which is:
\beqa
\label{eq:qpotau}
\tau_{k}\sim 2\pi\, \nu_{k}\,\frac{\,l_{k}^2\,\,}{ \,T_{{\rm eff}}\,}\frac{\delta\,\!M\,}{\delta\,\!V},
\hspace*{10mm}
\eeqa\\
where $\delta\,\!M$ is the mass involved in the crust oscillations, and $\delta\!\,V\,=\,\alpha \,l^3_{k}$ denotes the involved characteristic volume, where $\alpha$ is attributed for the shape of the volume. With given $B^2_{\rm{eff}}$, $\rho_{k}$, $l_{k}$, $\delta\,\!V$,
and $\delta M$, $\nu_k$ and $\tau_k$ can be solved with iteration through Equations \eqref{eq:qpofre} and \eqref{eq:qpotau}. If the whole crust/core is evolved in the oscillation, i.e., $\alpha\approx1$, we take $l_k\approx10^6$\,cm,  $\rho_{k}\approx10^{14}$\,g/cm$^3$, and $\delta M\approx3\times10^{31}$, $B_{\rm{eff}}\approx2\times10^{15}$\,G. For the mode $k=1$, the corresponding $\nu_{k}$ and $\tau_{k}$ are calculated to be $\nu_{k}\approx22-23$\,Hz and $\tau_{k}\approx0.2$\,s, respectively, which are in accordance with the observed QPO's 22.5\,Hz frequency and 0.2\,s duration.

\section{Summary and Discussion}
\label{sec:3} 
We proposed a new model to explain the observed precursor and its QPO in GRB 211211A. In our model, the progenitor of the GRB is a binary composed of a magnetar and an NS. At the late phase of inspiral, where the binary separation was $\sim90$ km, the tidal force on the magnetar, arisen from its NS companion, surpassed the maximum deforming force that its crust could sustain. The maximum sustainable deformation force was significantly reduced by the nonuniformity of its crustal magnetic field in the case of a magnetar. As a result, a global-scale crust destruction occurred and the magnetic field energy interior to the magnetar was released into the magnetosphere, causing the observed magnetar SF, as a precursor to the MB. The breaking crust, linking the magnetosphere and interior with strong magnetic field lines, oscillated under magnetic tension. The oscillating frequency corresponded to the observed QPO frequency of the precursor and disassociated within $\sim0.2$\,s, which explains the duration of the QPO.

\subsection{Correlation on Observable Properties Predicted from the Model}
From Fig. \ref{fig:Sfactor} we notice that the time interval between the beginning of the SF and the coalescence is positively correlated with its magnetic field strength. Since the energy of the flare is attributed to the magnetic energy, we therefore expect that a positive correlation between the time interval and its energy. For the same reason, the QPO frequency is also expected to be positively correlated with the total energy. However, we do not expect such correlations can be observed in the existing archival data, due to the small size of the sample of sGRBs with precursors. Besides, the energy of precursors of most sGRBs are poorly constrained, due to the lack of robust distance (redshift) information.

\subsection{The Compact Companion}
In the case that the magnetar has a normal NS companion, we expect $\rho_{\rm{eff}}$ of the NS to be $\sim3\!-\!10$ times larger than that of the magnetar (Fig.~\ref{fig:Sfactor}), due to less contribution to the deformation from the crust magnetic field. As a result, the crust creaking of the NS will happen a fraction of a second later than the magnetar. Since a normal NS is expected to have a magnetic field $1-3$ orders of magnitudes less than that of a magnetar, the energy released in the crustal destructive event of the normal NS would be less than $0.1\%-1\%$ of that from a magnetar. Therefore, such crust creaking from the NS companion is difficult to detect. When the companion is another magnetar, there could be two precursor peaks. Their temporal and spectral features will depend on the specific magnetic field configurations of the two magnetars. In the case of a black hole (BH) companion, the precursor can occur a few seconds prior to the merger. For a binary with a 10 $M_{\odot}$ BH and a magnetar with typical parameters as we used above, the precursor is estimated to happen $\sim1.5$ s before the merger.

\subsection{Implication for Short GRBs}
The tidal-induced crustal creaking mechanism is anticipated to be applicable in DNS mergers in general. Since sGRBs are widely believed to originate from DNS mergers, the nondetection of a precursor in some sGRBs can be used to place limit on the magnetic field strength of their progenitor NS. The observation of such a precursor SF indicates the presence of a magnetar in our framework. The fraction of sGRBs that have SF precursors thus provides an implication of the portion of DNS with at least one magnetar.
The fraction of the sGRBs with precursors is estimated to be $\lesssim1\%$ \citep{Minaev:2016gck,PhysRevD.102.103014}, which is broadly compatible with the currently known $\lesssim1\%$ portion of magnetars in all NSs \citep{doi:10.1146/annurev-astro-081915-023329}, despite of possible observational bias and incompleteness.

Currently, there are about $\sim20$ sGRBs found with a significant precursor that lasts $\sim0.05-0.2~{\rm s}$ and is separated from the MB by a quiescent period of $\sim0.1-1~{\rm s}$, e.g., GRB090510A, GRB100223A, GRB100827A, GRB141102A, GRB150604A, GRB150922A, GRB160804B, GRB181126A, which can be interpreted by our SF model. However, robust distance information is required to identify if the sGRB precursors are SFs. Additionally, some of these precursors, which release energy less than $\sim10^{47}$ erg, may not be produced from magnetar SFs.

Since the SF emission is likely unbeamed whereas the MB emission is beamed, it is naturally expected that such an SF is visible from a much larger solid angle than the MB, although with lower brightness. If these SFs are detected without the MB, they will assemble a special group of sGRBs, which may show distinctive features (e.g., the occurrence before a merger, QPOs, a relatively faint peak flux) from the traditional sGRBs. Note that the SF duration is likely short, although a longer SF duration is also possible, which demands extreme model parameters.

\subsection{Prospect of Multi-messenger Observation of Similar Events}
The major GW detectors like LIGO and Virgo were not in operation during GRB 211211A, so there lacks evidence from GWs for the compact object merger nature of this GRB, nor for whether the observed precursor really happened at the eve of the final merger. Moreover, the distance of GRB 211211A is estimated at 346.1 Mpc, which is also well out of the expected DNS detection range of LIGO/Virgo/KAGRA in the O4 period\footnote{\url{https://observing.docs.ligo.org/plan/}}. MBs of GRBs are highly collimated emissions. As a result, although it is expected that $\sim$10-20 DNSs may be detected with GWs in O4\footnote{Estimated with the GW Universe Toolbox \citep{toolbox}: \url{https://gw-universe.org}.}, there might only be a couple of them that will be found as GRBs. Therefore, the chance of finding a GW-associated GRB is still low in O4. However, in our model, the precursor comes from the premerger phase and is not collimated. Thus, we do expect a fraction of the GW-found DNSs in O4 to be associated with GRB 211211A-precursor-like $\gamma$-ray flares, which are seconds or tenths of second before the coalescence (assuming the velocity of GWs equals that of light). This fraction corresponds to the fraction of DNS systems that contain a magnetar. Such information provides unique clues to the evolution and population of progenitor binaries. The observation of such SFs, together with the information from the independent observation of GWs, e.g., the binary masses, tidal deformation parameters, orbital separation at the flare, etc., can bring us copious information on the structure of NS interiors, as well as the interior magnetic fields of the magnetars, which is hard to study from elsewhere. Such SFs will also be valuable in multimessengers/wavelengths follow-up observations, as they can play the role of earlier alarms and provide extra localization information.

\begin{acknowledgments}
We appreciate valuable comments and constructive suggestions from the anonymous reviewer that helped us improve the manuscript. This work is partially supported by the National Program on Key Research and Development Project (grant No. 2021YFA0718500) from the Minister of Science and Technology of China and the International Partnership Program of the Chinese Academy of Sciences (grant No.113111KYSB20190020). S.X.Y acknowledges the support by the Institute of High Energy Physics (grant No. E25155U1). S.L.X acknowledges the support by the Strategic Priority Research Program on Space Science of the Chinese Academy of Sciences (grant No. XDA15052700).
\end{acknowledgments}

\appendix

\section{Stressed crusts and energy extraction}  
\label{sec:1A}

Under the magnetic field of the magnetar, the crystals inside the crust are stretched nonuniformly. In reality, the crystal stretch is usually balanced by the anisotropic magnetic pressure. Thus, the crystals of the crust are already in an excited state. In other words, the crust is already stressed and its excited state incorporates the magnetic stresses. As the binary orbit shrinks, these crust crystals from one NS are continuously excited by the tidal force from the other NS or black hole. The crystals will be disrupted once the tidal force surpasses the breaking limit for the magnetized crust. 
After the crust is broken, the system is out of balance and, as a result, perturbations are produced in the form of plasma flows threading the core and former crust region.
This leads to the release of the huge magnetic energy reservoir stored in the core and crust through the propagation of the perturbations.
After the release of the interior magnetic energy and its conversion to particle energy and to radiation,
the crystals in crustal fragments settle into a new equilibrium state, which is less energetic and more relaxed.
In this process, the tidal force plays a key role in inducing the state transition in the late stage of a binary coalescence.

Note that there may be a discontinuity in the magnetic field between the core and crust due the presence of a current sheet caused by the complicated physics at the core-crust boundary \citep{Henriksson:2012qz}. It means that the magnetic field lines from the core can only penetrate a short distance into the crust. However, once the crust is tidally broken, the interior magnetic field is no longer isolated by the core-crust boundary. In other words, the magnetic field lines from the core can penetrate through the crust directly. If so, the original magnetic field inside the crust will be enhanced greatly; as shown in \cite{Lander2013}, the surface magnetic field strength can be of the same order as that at the core-crust boundary, which creates the condition for the effective release of the interior magnetic energy. Thus, the mechanism presented above still works.

As already mentioned, strong plasma perturbations can be caused in the crust and core by the crust breaking. These perturbations drive strong electric fields, which can accelerate charged particles, resulting in the huge release of the magnetic energy. In fact, the luminosity extracted from both the core and crust can be as high as \citep{Tsang2013}
\begin{eqnarray}
\label{eq:Lmax}
\nonumber
L_{\rm{max}}&=&\iint_{\Sigma}\(\vec{\upsilon}\times\vec{B}\)\times\vec{B}\cdot\d\vec{\Sigma}\\
&\sim&10^{51}~{\rm erg\,s^{-1}}\(\frac{\upsilon}{c}\)\,\(\frac{B_{\rm eff}}{\rm 10^{15} G}\)^2 \(\frac{R_{\Sigma}}{\rm 10\,km}\)^2,
\end{eqnarray}
where $\frac{\upsilon}{c}$ is the velocity of the perturbations in units of the speed of light, the radius $R_{\Sigma}$ characterizes the broken stellar surface $\Sigma$, and $B_{\rm eff}$ denotes the enhanced magnetic field strength at the radius of $r\sim R_{\Sigma}$ after the crust disruption. Here $B_{\rm eff}$ can be of the same order as the magnetic field strength at the outer boundary of the NS core~\citep{Lander2013}, and it can be 1-2 orders of magnitude higher than that at the magnetar surface before destruction~\citep{Henriksson:2012qz}. For instance, if the perturbations propagate in the same speed as Alfv$\acute{\rm{e}}$n waves, $\frac{\upsilon}{c}$ can be $\gtrsim0.1$ in the solid crust and the core where protons do not completely form a superfluid \citep{Levin:2006ck}. Thus, it is not hard to produce the observed SF luminosity, peaked at $7.4_{-0.7}^{+0.8}\times10^{49}~{\rm erg\,s^{-1}}$ \citep{Xiao:2022quv}. The total energy released in the SF precursor is about $7.7\times10^{48}$ erg, with a time-averaged luminosity of $\sim3.9\times10^{49}~{\rm erg\,s^{-1}}$ \citep{Xiao:2022quv}. Considering $E_{{\rm core}}\sim10^{50-51}$\,erg, which is much larger than the total energy release, such an SF precursor lasting $\sim0.2~{\rm s}$ does not have an energy budget problem. If the released energy propagates along or interacts with magnetic field lines, the resultant emission will be nonthermal. In addition, it can be expected that the DR mechanism continues to be at work after the crust breaking has been triggered so that more energy can be extracted effectively from the NS interior. In our model, the energy release and radiation mechanisms naturally lead to the unbeamed precursor flare.

\bibliography{draft}
\bibliographystyle{aasjournal}

\end{document}